# Multi-Fields Modulation of Physical Properties of Oxide Thin Films*

Huali Yang(杨华礼)[1,2], Baomin Wang(王保敏)[1,2,*], Xiaojian Zhu(朱小健)[1,2], Jie Shang(尚 杰)[1,2], Bin Chen(陈 斌)[1,2], and Run-Wei Li(李润伟)[1,2,†]

[1]*Key Laboratory of Magnetic Materials and Devices, Ningbo Institute of Materials Technology and Engineering, Chinese Academy of Sciences, Ningbo, Zhejiang, 315201, P. R. China*

[2]*Laboratory of Magnetic Materials and Application Technology, Ningbo Institute of Materials Technology and Engineering, Chinese Academy of Sciences, Ningbo 315201, P. R. China*

Correspondence and requests for materials should be addressed to
B. W. (wangbaomin@nimte.ac.cn) or R.-W. L. (runweili@nimte.ac.cn)

PCAS: 73.23.-b; 73.50.Pz; 77.55.Nv.

*Project supported by the State Key Project of Fundamental Research of China (973 Program, 2012CB933004), National Natural Science Foundation of China (11274321, 51301191, 61328402, 11274322), Overseas, Hong Kong& Macao Scholars Collaborated Researching Fund (51428201)，the Instrument Developing Project of the Chinese Academy of Sciences (YZ201327), Ningbo Major Project for Science and Technology (2014B11011), Ningbo International Cooperation Projects (2012D10018).

**Abstract**

Oxide thin films exhibit versatile physical properties such as magnetism, ferroelectricity, piezoelectricity, metal-insulator transition (MIT), multiferroicity, colossal magnetoresistivity, switchable resistivity, etc. More importantly, the exhibited multifunctionality could be tuned by various external fields, which has enabled demonstration of novel electronic devices. In this article, recent studies of the multi-fields modulation of physical properties in oxide thin films have been reviewed. Some of the key issues and prospects about this field are also addressed.

**Contents**

## 1. Introduction

Oxides present a class of materials that are particularly attractive due to their remarkable physical properties and multifunctionality.[1-4] For example, a diversity of physical properties, such as magnetism, ferroelectricity, piezoelectricity, and superconductivity, etc., exhibited in thin film oxides, have been intensively studied in the past several decades,[5-10] and the interplay between them has become a hot topic in condensed matter physics.[11-15] Intriguing phenomena including MIT,[16-24] colossal magnetoresistivity,[25-29] resistive switching,[30-38] ferroelectricity,[39-45] and multiferroicity,[46-56] etc., make oxides appealing candidates for both electronic and spintronic devices.[57-63] From fundamental perspective, the multifunctionality in oxides provides a playground for investigating the interplay between spin, charge, orbital and lattice degrees of freedom.[24,64-67] A number of interesting phenomena such as charge order, orbital order, multiferroicity, and interactions such as electron correlation effect, Jahn-Teller interaction, and spin-orbit coupling, has been frequently discussed to understand the fundamental physics.[51,64,68-73] In addition, the advent of atomically tailored oxide heterostructures has enabled the observation of emergent phenomena, and raised excitement in study of novel phenomena due to the symmetry breaking and charge transfer across the oxide interface.[67,73-76]

The interplay between spin, charge, orbital and lattice degrees of freedom in oxides (especially complex oxides) reveals strong composition-structure-property relationships, and enables the possibility for studying the multi-fields effect on the physical properties of thin film oxides. As is shown in Fig.1, the interconnections between light, magnetism, heat, strain, and electric has been demonstrated in condensed matters, and many novel phenomena such as photorestriction,[77] thermal spin-transfer torque,[78] and multiferroicity has been revealed in functional oxides. Many novel devices with multifunctionality have been proposed, such as spin-resistive random access memories (RRAM), electric operational magnetic random access memories (MRAM), multiferroic tunnel junctions, etc.

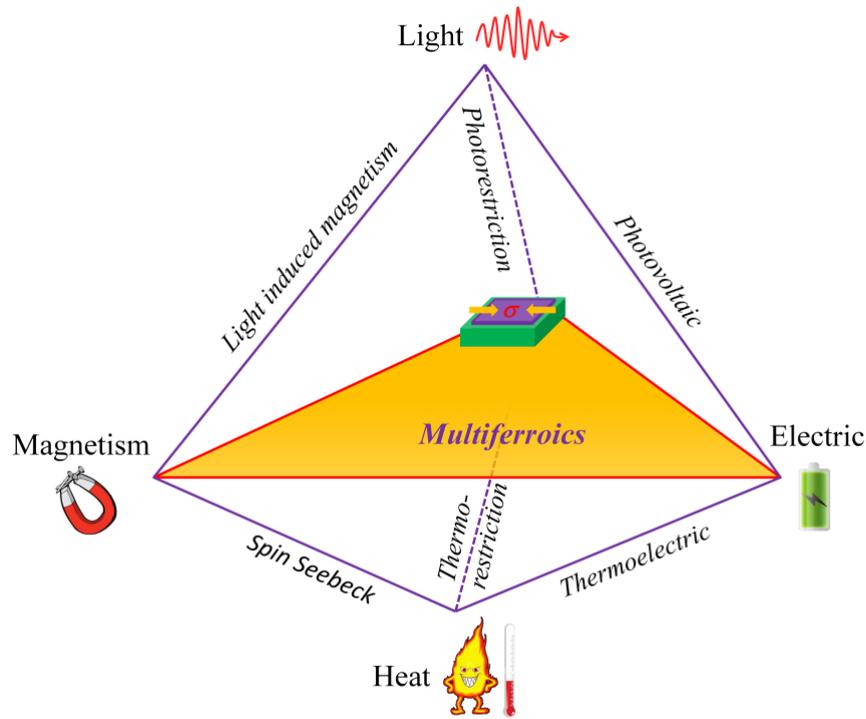

**Fig. 1** Schematic showing of the multi-fields (light, magnetism, heat, electric, and strain) modulation of the physical properties in oxides.

In many cases, combined external stimuli could enable fine tuning of physical properties in thin film oxides, which could be beneficial for practical applications. In the following paragraphs, we will discuss the response of thin film oxides upon the external stimulus such as electrical field, strain field, magnetic field and photo illumination. A special concern has been focused on multiferroics, perovskite manganites, and oxides with resistive switching (RS) properties.

## 2. Electric field effect

Electric field has been recognized as an efficient and scalable way for tuning the physical properties in various materials and structures. For example, in magnetic tunneling junctions, a spin polarized current could induce spin transfer torque effect, which has been intensively studied in MRAM devices[79-82] and microwave devices.[83] Upon the application of electric field, the polarization direction of ferroelectrics could be switched. This could result in either large electroresistance effect in ferroelectric tunnel junctions[84,85] or switchable diode effect in ferroelectric diodes.[86] Noticeable strain effect associates with the electric-field-induced ferroelastic domain switching has also been intensively studied in magnetoelectric heterostructures.[87,88] In multiferroic materials where the electric polarization couples with the magnetization, applying an electric field modifies the electric polarization and tunes its magnetization through the magnetoelectrical (ME) coupling effect.[11,89,90] In RRAM devices, applying an electric field induces the RS effect by associating with several electric-field-induced effects: electric field induced ion transportation,[35] electric field induced phase transition,[91] and the Joule heating effect.[92]

**2.1 Electric field control of magnetism**

The interplay between electric polarization and magnetization has attracted wide interest for applications such as spintronics devices.[48,93] The presence of multiferroicity in some materials and structures has enabled several ways where the magnetic properties can be manipulated by applying an electric field:[94] (1) In single phase multiferroics where magnetic order and ferroelectricity couples, applying an electric field could alter the electric polarization along with the magnetization owing to the intrinsic ME coupling.[11] (2) Many multiferroics are antiferromagnets, by attaching it to a ferromagnetic layer, it is possible to construct an exchange bias system that could be controlled by applying electric field to the multiferroic layer, which is known as exchange-mediated ME coupling effect[12]. (3) In magnetoelectric composites composed of ferroelectric (FE)/ferromagnetic (FM) heterostructures,[95] applying an electric field in the FE layer would induce strain that are transmitted in to the FM layer and changes its magnetic properties through inverse magnetostrictive effect. (4) In FE/FM heterostructures, the electric field could also switch the polarization direction of the FE layer, which may tunes the carrier density and hence the magnetic properties of the FM layer, referring as charge-mediated ME coupling effect.[96]

In single phase multiferroic materials, a tempting idea is to design multiferroics that combine both ferromagnetism and ferroelectricity, and the manipulation of its magnetization by an electric field would be essential for the magnetoelectric control of spintronics devices. However, the exploring of both ferromagnetism and ferroelectricity in single phase multiferroic has been found unexpectedly difficult.[46] One promising single phase multiferroic that could be utilized is $BiFeO_3$ (BFO) — a material that shows unambiguously ferroelectricity and anti-ferromagnetism above room temperature.[97] The principle that BFO can be utilized for an efficient magnetoelectric control of spintronics devices relies on the possibility of electric field controlled interfacial coupling that composed of BFO and a ferromagnetic layer.[12,98,99]

Recently, Heron *et al.* demonstrated the deterministic magnetization reversal in BFO based heterostructures,[99] by utilizing the coupling between its weak magnetization arising from the Dzyaloshinskii-Moriya (DM) interaction and the ferroelectric polarization,[100,101] as well as the strong interfacial magnetic coupling.[102] Although previous theory[103] predicts that the direct reversal of ferroelectric polarization in BFO will leave the orientation of the canted magnetic moment unchanged, they found that ferroelectric-polarization reversal in a strained BFO film actually follows an indirect pathway, and proposed that the indirect switching path reverses the canted magnetic moment [Fig. 2]. The ME coupling is determined by the DM interaction, where the weak ferromagnetism $M_c$ is determined by the DM vector which is given by $D \sim \Sigma d_i \times r_i$, where $r_i$ is a constant vector connects neighboring $Fe^{3+}$ ions and $d_i$ is the displacement of the intermediate oxygen atom from the mid-point of $r_i$ caused by the $O_6$-rotation. Since the in-plane 71 ° and out-of-plane 109 ° switches are found to be capable of rotating $D$ and the associated $M_c$ by 90 ° [Fig. 2(b) and 2(c)], the reversal of $D$ through sequential in-plane 71 ° and out-of-plane 109 ° switches would also be expected to reverse $M_c$. Furthermore, they deposited a spin valve structure on top of the strained BFO, and realized the electric-field control of a spin-valve device at room temperature.

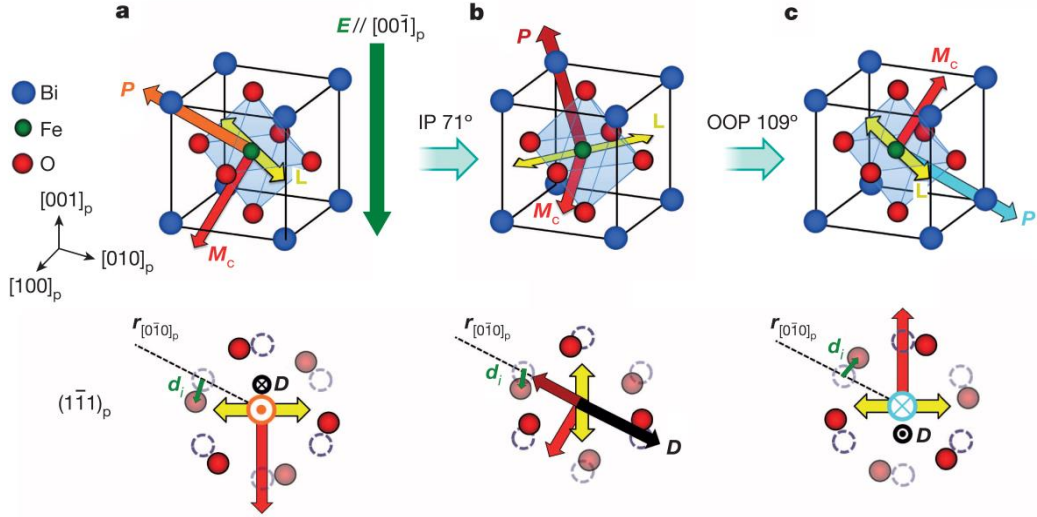

**Fig. 2** (a) A domain with polarization ***P*** initially along the $(1\bar{1}1)$. ***L*** represents the antiferromagnetic axis. $O_6$ rotations are shown with respect to an octahedron without rotations (dashed circles). (b) After an in-plane 71 ° switch. ***E*** represents the applied electric field. (c) After an out-of-plane 109 ° switch.[99]

The progresses of strain-mediated and charge-mediated ME effect will be further discussed in later sections.

**2.2 Electric field induced ion migration**

RRAM has been widely investigated as next-generation nonvolatile memory devices due to its simple structure, high density, low power consumption and compatibility with complementary metal oxide semiconductor technology.[61,104-106] Up to now, a large variety of materials have been investigated for RS including solid electrolytes,[107] perovskite oxides,[32,108-110] binary oxides,[33,35,37,111] organic materials,[112-114] amorphous silicon,[115] and amorphous carbon.[116] Although the underlying mechanism is still controversial due to the diversity of RS behaviors, conductive filament (CF) model has gained much success and has been accepted extensively.[34]

RRAM devices show a simple metal-insulator-metal sandwiched structure. For filament model based RRAMs, a sufficiently large electrical field between the two metallic electrodes could induce a soft dielectric breakdown in the insulator, and set the device to a conductive state (also refer as an ON-state). In this process, due to the electric induced ion migration, a large amount of defects are formed inside the film and form CFs that connect the two terminal electrodes. The ON-state could also be electrically reset, leading to the OFF-state that corresponds to disruption of the CFs.

In the CF model, the formation and rupture of local conducting paths connecting the two-terminal electrodes dominates the RS process.[36,38,117,118] Liu *et al*. have performed real-time characterizations of the CF formation and dissolution processes of an oxide-electrolyte-based RRAM device,[36] which is based on a vertical Ag(Cu)/$ZrO_2$/Pt structure with both small lateral sizes and $ZrO_2$ thickness to facilitate direct TEM studies. During the SET process, a dark nano-bridge region appears, signifying the formation of CF [Fig. 3(a)]. Successive RESET process has been performed, and a first RESET voltage sweep is found to decrease the electrical current passing through the device drastically [Fig. 3(b)]. Meanwhile, the CF region near the Pt electrode was substantially weakened after the first RESET process [Fig. 3(c) and 3(d)]. The chemical compositions of the CF region in the initial-, ON- and OFF-states were analyzed using energy-dispersive x-ray spectroscopy (EDX) analysis and are found to be Ag element in the Ag/$ZrO_2$/Pt device [Fig. 3(e)]. These experiments have led to intuitive and more complete understanding of the underlying nature of the RS behavior and the corresponding CF evolution process.

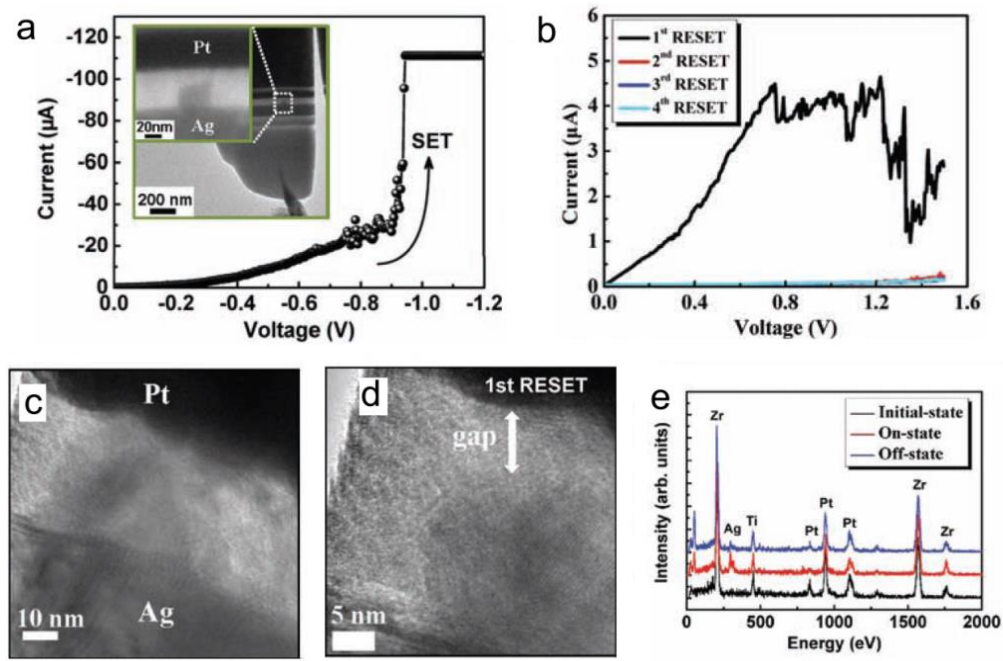

**Fig. 3** (a) *I-V* curve of the Ag/ZrO$_2$/Pt TEM specimen during SET operation. The inset shows the TEM image of the specimen after the SET process. (b) *I-V* curves of the TEM specimen under successive RESET operations. (c) High resolution TEM image of the CF region in an Ag/ZrO$_2$/Pt TEM specimen after SET process. (d) High resolution TEM images showing that the CF dissolves from the anode (Pt) to the cathode (Ag) terminal during RESET voltage sweeps (0 →1.5 V). (e) EDX analysis of the TEM specimen conducted at the initial-, ON- and OFF-states.[36]

It has been found that the nanoscale conduction channel not only dominates the RS process but also shows some other effects, such as quantum size effects and exchange bias effect. For example, quantized conductance has been realized in zinc-oxide based sandwich devices.[35] As can be seen from Fig. 4, the Nb/ZnO/Pt device shown in Fig. 4a can be Set and Reset repeatedly, exhibiting typical bipolar RS behaviors [Fig. 4(b)]. During the Set process where the nanoscale conducting channel is forming, the conductance shows quantum behavior corresponding to electron travelling along the nanoscale channel [Fig. 4(c) and (d)]. Similar phenomena have also been found in other sandwiched thin films.[118-122]

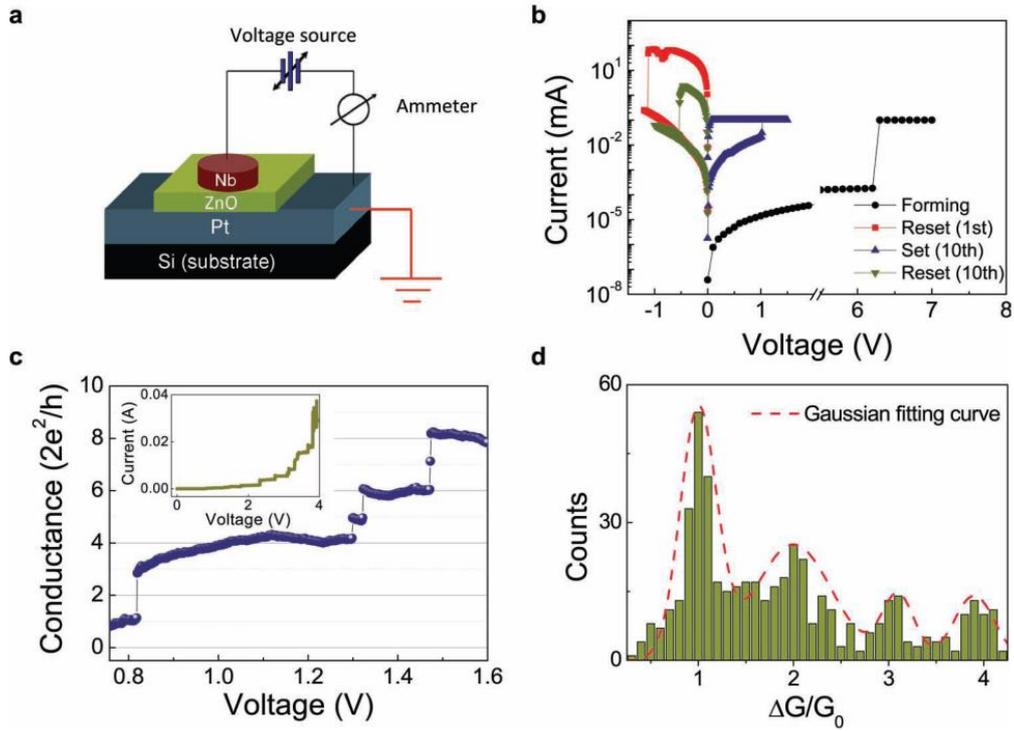

**Fig. 4** (a) Schematic of a sandwiched Nb/ZnO/Pt structure. (b) Typical bipolar RS behaviors observed in the device. (c) Measured conductance as a function of the bias voltage during the Set process. Inset: Corresponding I-V curve in a larger voltage range from 0 to 4V. (d) Histogram of the conductance changes ΔG obtained from the Forming and Set processes.[35]

It has also been revealed that by controlling the migration of ions through electrical fields, the magnetism of some magnetic dielectrics can be modulated. For example, Chen *et al*. have successfully modulated the magnetization and coercive field of Co-doped ZnO thin films.[123] Similar behavior has been reported in other materials such as Mn:ZnO[124] and $Fe_3O_4$.[125] In $CoFe_2O_4$ (CFO) thin films, nanoscale magnetization reversal *via* electrical ion-manipulation at room temperature has been reported.[126] As is shown in Fig. 5(a), after subjected to various bias voltages, the MFM signal of CFO thin films indicates magnetization reversal. The electrical modulates the magnetization in CFO thin films in a reversible and non-volatile manner [Fig. 5(b)], and can be realized without the assistance of external magnetic field. This strategy of utilizing electric field-induced migration and redistribution of ionic species to modulate the nanoscale magnetization is favorable for the construction of novel spintronic devices.

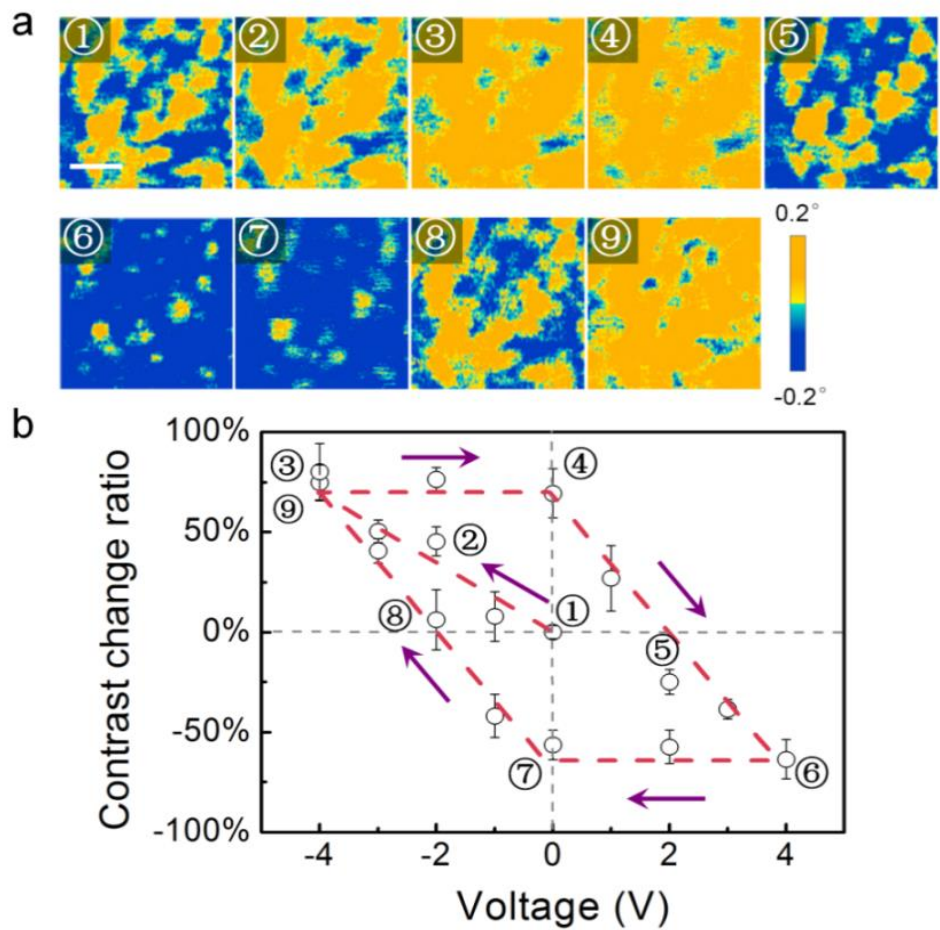

**Fig. 5** (a) MFM images of the CFO thin film at the pristine state and after being subjected to various biased voltages. The scale bar is 200 nm. (b) Evolution of the contrast change ratio (defined as the ratio between the total areas of the electrically-produced upward-magnetized domains and the initially upward-magnetized domains) after being subjected to various biased voltages in the sequence of 0 V → -4 V → 0 V→ 4 V →0 V → -4 V.[126]

**2.3 Electric field induced MIT**

In thin film oxides, besides ionic transportation process, the RS phenomenon is associated with other mechanisms such as self-Joule heating effect[92] and electric field induced electronic phase transition.[91] The electric field induced electronic phase transition has been predicted theoretically in charge-orbital ordered manganites, where applying an electric field may suppress the charge ordering and drive the system from the antiferromagnetic charge-ordered state to the ferromagnetic metallic state, resulting in the well-known MIT.[127] The electric field induced magnetism in manganites shows a new way of realizing the electrical control over magnetism.[128,129] Another very interesting and widely studied correlated material with electric field induced MIT is vanadium dioxide ($VO_2$). Bulk $VO_2$ exhibits MIT at temperature ~340 K.[130] The electric field was found to be able to drive MIT in $VO_2$, accompanying with a structural change[131] with large isothermal entropy change.[132] Various electronic devices have been proposed based on the electric field induced MIT in $VO_2$.[133,134] However, the underlying mechanisms have not been fully clarified.[135-138]

**2.4 Electrical modulation of the carrier density**

The electronic structure of some functional oxides (such as manganites) depend critically on the doping level, hence enables electrostatic field tuning of their physical properties.[139,140] The charge-mediated tuning of functionality (such as electric transportation and magnetization) in oxides has been studied both theoretically[141-144] and experimentally[96,145,146] in a number of heterostructures.

In perovskite manganites, the electrostatic field could modify the charge carrier concentration without affecting the lattice distortion, which is useful to study of the influence of band filling on the electronic properties. The only challenge comes from the high carrier concentration in FM manganites (with carrier density as high as $10^{21}$ carriers per $cm^3$) that result in an electronic screening length on the order of only 1 nm (the Thomas-Fermi screening limit), which has limited the tunability of electrostatic field imposed by utilizing conventional dielectrics like $SiO_2$, $SrTiO_3$ (STO), and ferroelectrics such as lead zirconate titanate (PZT).[140,147] In recent years, the electric double-layer transistor (EDLT) has been used to achieve high electric fields, enabling charge accumulation in the channel surface at densities as high as $10^{15}$ per $cm^2$.[148] This is almost two orders of magnitude larger than that in conventional field effect transitions, thus providing a more accessible way to the electrostatic modulation of electronic phases. The effective tunability by using EDLT has been demonstrated in manganite transistors such as (La,Ca)MnO$_3$ (LCMO) transistors,[149] (Pr,Sr)MnO$_3$ (PSMO) thin films,[150] and (Pr,Ca,Sr)MnO$_3$ (PCSMO) thin films.[151] For example, in PCSMO thin films, it is found that, in consistency with phase diagram, the electrostatic modulation of the carrier concentration could effectively modify the phase transition temperature. The electrostatic modulation could also co-play with external magnetic field, resulting in the fine tuning of electronic phase and phase transitions.[150,151]

A MIT can be induced by a gating field in VO$_2$, and the electrostatic effect caused by the gating field was found by Nakano *et al.* to have overcome the Thomas–Fermi screening limit.[152] They suggested that the Thomas–Fermi screening limit does not apply to materials such as VO$_2$, which is characterized by inherent collective interactions between electrons and the crystal lattice. Jeong *et al.*, however, showed that the electrolyte gating of VO$_2$ leads not to electrostatically induced carriers but instead to the electric field induced creation of oxygen vacancies. They found that by sweeping V$_G$, the device could be reversibly switched between the insulating- and metallic- states. Hysteresis in the sheet conductance centered about V$_G$ = 0 V was observed [Fig. 6(a)], and the high conductance state maintained for many days even when the ionic liquids was washed off. XPS was taken to measure changes in the oxidation state of vanadium in VO$_2$ films, and they have observed a reduction in the oxidation state of V from V$^{4+}$ toward V$^{3+}$ when the film is gated [Fig. 6(b)], indicating migration of oxygen from the oxide film into the ionic liquid. These results are in consistence with the correlation between the source-drain current and the amount of oxygen pressure [Fig. 6(c)].

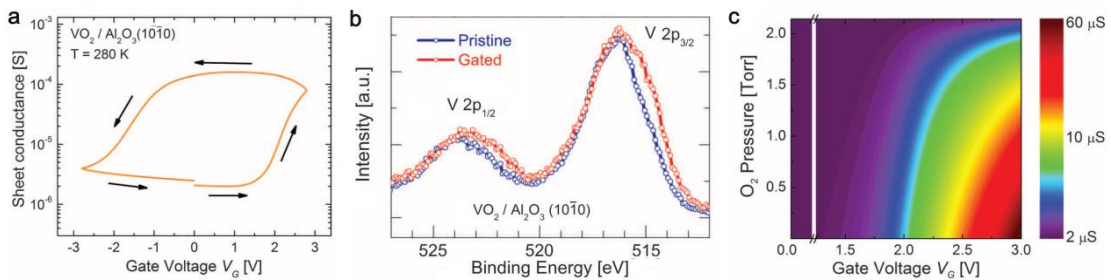

**Fig. 6** (a) Sheet conductance versus $V_G$ for devices fabricated from $VO_2$ films prepared on $Al_2O_3$(1010). (b) V 2p core-level spectra for pristine and gated $VO_2/Al_2O_3$ (1010) at $V_G$ =1.8 V. (c) Sheet conductance (color scale) as a function of $V_G$ and $O_2$ pressure.[153]

The mechanism associates with the electric field induced oxygen migration has also been suggested by Cui *et al.*,[154,155] in the ionic liquid gating experiments on (La,Sr)MnO$_3$ (LSMO). Although the oxygen vacancies migration has not been directly observed, the Mn valence variation in XPS and microstructure evolution in HRTEM under different gate voltages suggest the formation and annihilation of oxygen vacancies. The electric fields created by the electric double layer (EDL) on a manganite under $V_G$ will change the chemical composition, which is accompanied by changes in the effective doping level *x* [Fig. 7(a)]. Interestingly, an effective tuning of the orbital occupancy and corresponding magnetic anisotropy by gate voltage have been observed by using x-ray linear dichroism (XLD) [Fig. 7(b) and 7(c)]. The tuning behaves in a reversible and quantitative manner, manifesting advances towards practical oxide-electronics based on orbital.

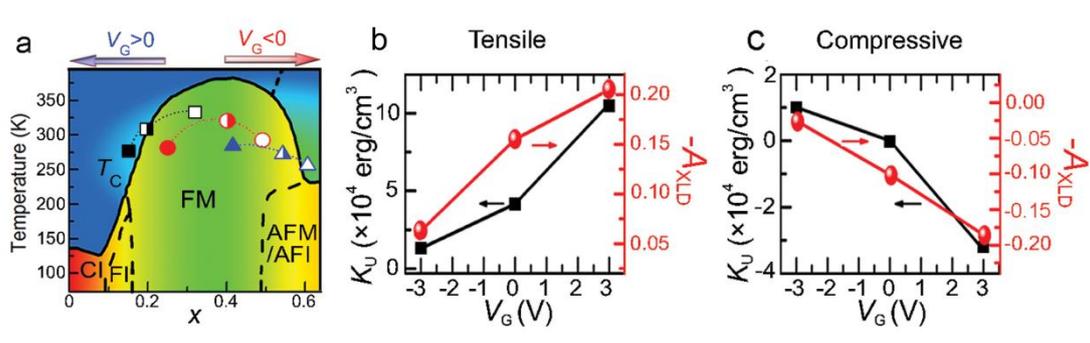

**Fig. 7** (a) Electronic phase diagram of LSMO with varying $x = Mn^{4+}/(Mn^{3+} + Mn^{4+})$ ratio. The half-filled triangle, circle, and square denote the positions of samples on STO with initial *x* = 0.54, 0.41, and 0.20 in phase diagram, respectively. The sample positions after applying positive and negative *V*$_G$ are marked by corresponding filled and empty symbols. The abbreviations stand for spin-canted insulating (CI), ferromagnetic insulating (FI), ferromagnetic metal (FM), antiferromagnetic metal (AFM), and antiferromagnetic insulating (AFI) phase, respectively. The dependence of $K_U$ (left axis) and $-A_{XLD}$ (right axis) on $V_G$ for the case of (b) tensile and (c) compressive strain.[155]

## 3. Strain effect

Applying strain on thin film oxides has been proved an efficient way for tuning their physical properties. Exerting strain on thin films could induce distortions of the film unit cell, which changes the elastic energy and modifies their electro-magnetic properties.[156] In some cases, applying strain could alter the ground state of thin film oxides and novel phases are disclosed. There are three ways to apply strain in thin films: mechanical strain,[157] epitaxial strain,[158] and strain that due to the converse piezoelectric effect.[159] In recent years, inspired by the idea of wearable electronic devices, there has been increasing interest in the study of flexible electronic devices,[160-164] and the strain effect in these devices has been intensively investigated.

### 3.1 Strain tuning of the electronic phases

In some transitional metal oxides, the electronic phases are very sensitive to external strain,[165,166] resulting in highly sensitive electrical and magnetic response to external strain field. For example, the strain effect on the electrical properties of manganites thin films has been studied by using the converse piezoelectric effect, the resistance of manganite thin films is found to change dramatically under small strain, resulting in very large strain gauge factor.[159,167-169] The electrical and magnetic response to external strain is essentially mediated by the piezostrain coupled orbital reoccupation[170] and electronic phase separation.[169] Zheng *et al*. investigated the influences of piezostrain on the electronic transport and magnetic properties of charge-ordered $La_{0.5}Ca_{0.5}MnO_3$ thin films.[169] An unprecedented large strain-induced resistance change is found to be magnetic-field tunable [Fig. 8(a)], indicating that this coupling is essentially mediated by the electronic phase separation. By using the ferroelastic domain switching of the ferroelectric substrate [Fig. 8(c) and 8(d)], they found that an in-plane tensile strain (correspond to the $P_r^{//}$ state) as small as 0.1% according to x-ray diffraction [Fig. 8(e) and 8(f)] could induce a large increase in the resistance and $T_{CO}$ and a drop in magnetoresistance, signaling the stabilizing of the charge-ordered phase. On the other hand, an in-plane compressive strain (correspond to the $P_r^-$ state) suppresses the charge-ordered phase. By taking advantage of the ferroelastic strain, an electric-field-controlled non-volatile RS behavior could be achieved [inset of Fig. 8(a) and Fig. 8(b)]. Besides the large strain induced resistivity change, applying anisotropic strain field is found to induce highly anisotropic resistivity in perovskite manganites.[171,172]

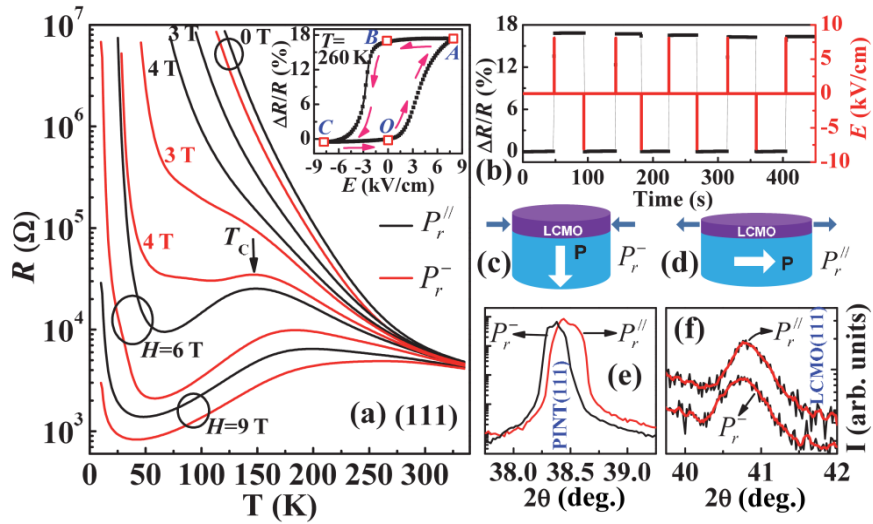

**Fig. 8** (a) Temperature dependence of resistance for the LCMO(111) film in the magnetic fields as stated when the ferroelectric substrate was in the $P_r^{//}$ and $P_r^-$ states, respectively. Inset in (a): $\Delta R/R$ of the film as a function of $E$ at $T = 260$ K. (b) Nonvolatile resistance switching of the LCMO film by a pulse electric field at $T = 260$ K. (c) and (d) Schematic diagrams for the ferroelastic strain effect at $T = 260$ K. (e) and (f) XRD $\theta$-$2\theta$ scans for the ferroelectric substrate and LCMO film when the ferroelectric substrate was in the $P_r^{//}$ and $P_r^-$ states, respectively.[169]

The substrate strain could sometimes induce new phases with dramatic electronic transportation properties. For example, it has been predicted theoretically that in half-doped manganite, the ferromagnetic charge ordered insulating state, previously inaccessible to experiments, becomes stable under tensile strain.[173] Recent report on LSMO thin films deposited on DyScO$_3$ (DSO) substrate has evidenced a new phase at low temperature with dramatic anisotropic transportation properties [Fig. 9(c), 9(d) and 9(e)].[158] The new phase shows preferred occupancy of the Mn $3d_{x^2-y^2}$ orbital at low temperature, which is consistent with the theoretically predicted A-type AFM phase [Fig. 9(a)].[165] However, abnormal in-plane transport anisotropy is observed in this low-temperature phase [Fig. 9(b) and 9(c)]. Note that only the film on DSO under large tensile strain shows pronounced resistivity anisotropy [Fig. 9(d)]. A switching of anisotropy sign from negative to positive upon cooling is also observed. The observed anisotropy is suggested to be driven by the preferred occupancy of the O $2p_x$ orbitals, which hybridizes with Mn $3d$ orbitals.

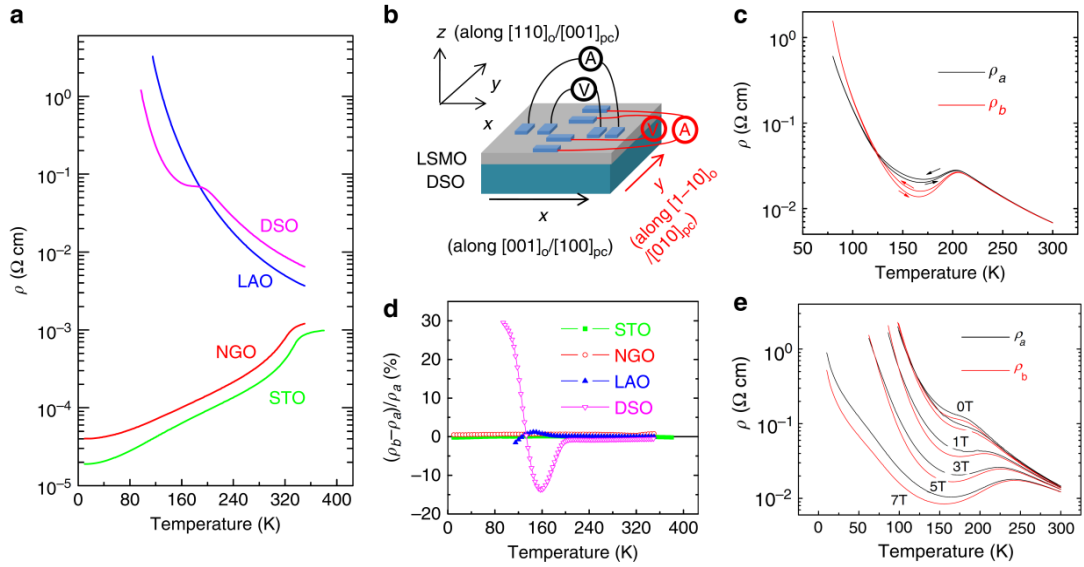

**Fig. 9** (a) Temperature dependence of resistivity of 8 nm LSMO thin films on different substrates under zero magnetic field. (b) Schematic illustration of the experimental set-up used to measure the in-plane resistivity of the LSMO thin films along the two orthogonal directions (*x* and *y*). (c) Resistivity versus temperature curves for 9.6nm LSMO on DSO along the two channels during cooling and heating processes. (d) The in-plane resistivity anisotropy, ($\rho_b - \rho_a$)/$\rho_a$, of 8nm LSMO thin films on different substrates. (e) The resistivity versus temperature curves under different magnetic fields for 9.6nm LSMO films on DSO.[158]

## 3.2 Strain control of ferroelectric domain structures

In ferroelectric materials, the formation of ferroelectric domain structures is a consequence of minimizing the elastic and electrostatic energy of a ferroelectric system. Hence applying strain can effectively tune the ferroelectric domain structures, which could be useful for further exploring the ME properties in multiferroic systems. Both theoretical and experimental studies have been performed on controlling the ferroelectric domain structures of BFO thin films, and strain constraints have been found to be an effective way for controlling the domain structures of BFO thin films. Zhang *et al*.[40] try to understand how the strain state can affect the polarization variants and to predict the domain structures in epitaxial BFO thin films with different orientations. These findings suggest that the domain structures of BFO thin films can be controlled by selecting the proper film orientations and strain constraints. Chu *et al*.[174] have demonstrated the realization of one-dimensional nanoscale arrays of domain walls in epitaxial BFO/ SrRuO$_3$ (SRO)/ DSO structures. This quasi-periodic structure results from the strain induced by the anisotropic in-plane lattice parameters of DSO ($a_1$ = 3.951 and $a_2$ = 3.946 Å), which pins the structure of the SRO layer and, eventually, the ferroelectric domain structure of BFO. Through the careful control of electrostatic boundary conditions, such as the thickness of the underlying SRO layer, Chu *et al*.[175] demonstrated the creation of ordered arrays of the prototypical domain structures predicted before.[176] The selection of 71° or 109° domain patterns is dependent on the electrostatic boundary conditions, such as the existence of a SRO conducting layer and its thickness. When the bottom electrode SRO layer is thick enough to be a good metal, the elastic energy is dominating in the system, and the domain is fully out-of-plane polarized downward towards the SRO layer, forming ordered arrays of 71 ° domain walls. On the other hand, when the SRO layer is very thin, electrostatic energy becomes the dominant energy and the domains are alternatively pointing up and down, with arrays of 109 ° domain walls favored.

Piezoelectric materials with large piezoelectric coefficients (such as lead-based perovskites PZT, PMN-PT, and PZN-PT) are typically characterized by the intimate coexistence of two phases across a morphotropic phase boundary,[177] and an electrically switching from one phase to the other.[178] By using a combination of epitaxial growth techniques in conjunction with theoretical approaches, Zeches *et al.* has shown the formation of a morphotropic phase boundary through epitaxial constraint in lead-free piezoelectric BFO films.[41] The tetragonal-like phase can be reversibly converted into a rhombohedral-like phase, making this new lead-free system of interest for probe-based data storage and actuator applications. Ab initio calculations demonstrated the strain-induced structural change in BFO [Fig. 10(a) and 10(b)] can be realized at a certain value of epitaxial strain provided the absence of misfit accommodation through dislocation formation. Direct atomic resolution images of the two phases [Fig. 10(c) and 10(d)] clearly show the difference in the crystal structures. For films grown at intermediate strain levels (e.g., ~4.5% compressive strain, corresponding to growth on $LaAlO_3$ substrates), a nanoscale mixed-phase structure [Fig. 10(e) and 10(f)] has been observed. Figure 10g is an atomic resolution TEM image of the interface between these two phases and reveals one of the most provocative aspects of these structures.

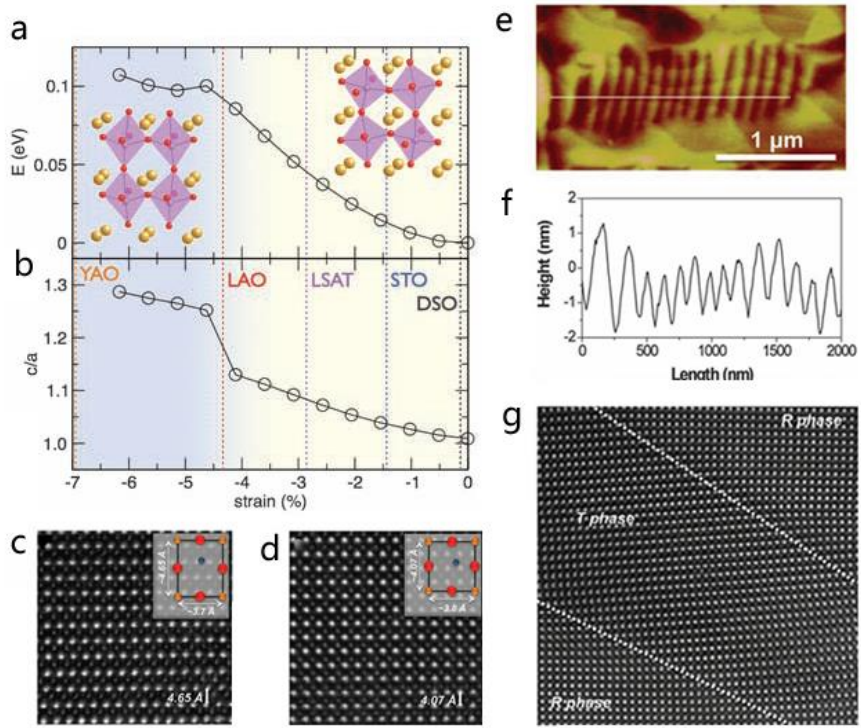

**Fig. 10** First-principle calculations provide information on the strain evolution of (a) the overall energy of the system and (b) the c/a lattice parameter ratio. High-resolution transmission electron microscopy (HRTEM) reveals the presence of two phases (c) a high-distorted monoclinic version of a tetragonal structure and a monoclinic version of the bulk rhombohedral phase (d). These complex phase boundaries manifest themselves on the surface of the sample as imaged via (e) atomic force microscopy and these features correspond to dramatic surface height changes as shown from (f) the line trace. (g) HRTEM imaging of boundaries shows a smooth transition between phases.[41]

**3.3 Strain mediated electrical control over magnetism**

A number of works have been devoted to the studying of composite multiferroic films,[96,179,180] most of which are composed of ferroelectrics and magnetic materials with layered heterostructures. In these composite multiferroic films, changes in magnetic anisotropy and coercivity can be realized by applying a strain,[181,182] which is caused by applying an electric field to the piezoelectric layer. The strain-mediated electric field control of magnetism has been observed in a number of epitaxial heterostructures, such as ferromagnetic LSMO films,[183] ferromagnetic CFO films[184] and metals grown on PMN-PT single crystal substrate, and ferromagnetic LSMO films[95] and metals[185] grown on BTO substrate. A giant electrical-field-induced magnetic anisotropy has been evidenced in these layered multiferroic heterostructures, and the strong ME coupling is used to control the orientation of magnetization and thus enable dynamically tuning of magnetoresistance in giant magnetoresistance (GMR) and anisotropic magnetoresistance (AMR) devices.[186,187]

In most strain mediated electrical control over magnetism case, the M-E loop is butterfly-shaped, tracking the butterfly-shaped piezo-strain curve of the ferroelectric substrate, which clearly demonstrates the vital part of the elastic strains for the converse ME response in the heterostructure. For the application of information storage, tunable and nonvolatile converse ME effects are two important requirements, hence loop-like behavior of the magnetic response to electric field is desirable. Therefore great efforts have been devoted to the non-volatile control over magnetism.[88,188-192] For ferroelectric materials with compositions near the morphotropic phase boundary, the electric-field-induced phase transitions are very prominent. For example, a rhombohedral-to-orthorhombic phase transition takes place in (011)-oriented PZN(6-7%)-PT under sufficient poling fields. Most of such phase transitions are non-volatile, and it is expected to display a hysteresis-type lattice change as a function of electrical-field. This effect can be used for realizing non-volatile spintronics and microwave devices based on multiferroic heterostructures.[180,190]

Yang *et al*. reported piezostrain-mediated nonvolatile rotations of magnetic easy axis (MEA) in the Co/PMN-PT multiferroic heterostructure at room temperature.[193] They found a reversible and non-volatile 90°rotation of magnetization [Fig. 11A (b) to (j)], which was induced by the asymmetric butterfly-like piezostrain-electric field loop for PMN-PT [Fig. 11A (a)], as has been understood by means of phase-field simulations [Fig. 11B (stage 1 and 2)]. Unexpectedly, a non-volatile 180° magnetization reversal is achieved without an applied magnetic field [Fig. 11B (stage 3)], and this unexpected 180° magnetization reversal, as is shown by phase field simulation, is mediated by simultaneous action of the piezostrain and perturbation field $H_{per}$ that is induced by the polarization switching current. A reversible 180° magnetization reversal can also be realized with an about 5 Oe auxiliary magnetic field [Fig. 11B, from stage 4 to stage 1]. Their discovery of the electric field-induced non-volatile, repeatable magnetization rotation could be helpful in designing electromagnetic devices.

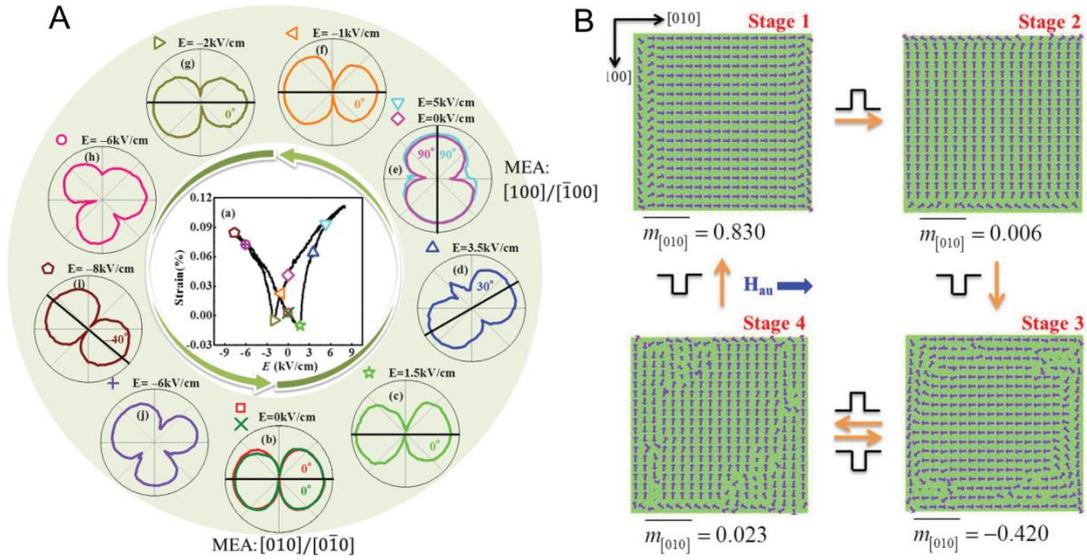

**Fig. 11** A. (a) In-plane strain ($\varepsilon_{11}$)–electric field loop along the [100] direction of PMN-PT. b–j) The rotation history of the MEA in the Co film with a sequence of electric fields (labeled by corresponding symbols on the strain curve) applied to PMN-PT along the [100] direction in order from (b) to (j). In (e), after the field of 5 kV cm$^{-1}$ is switched off, the MEA does not go back to 0° but remains at 90°. B. The magnetization domain at stages 1–4 under a stimuli sequence of ±5 kV cm$^{-1}$ (up and down pulses denoting respectively +5 kV cm$^{-1}$ and −5 kV cm$^{-1}$) accompanied by the $H_{per}$ of 2 Oe induced by the polarization switching current. An auxiliary field $H_{au}$ is used to return the magnetization to stage 1 from stage 4.[193]

Another way of realizing non-volatile electric-field-controlled magnetization is by inducing two different remnant strain states in the single crystal piezoelectric substrate.[194] However, such a way is not practical for real applications, as the remnant strain states strongly depends on the history of applied electric fields.

In the (001)-oriented PMN-PT substrate, a bipolar loop-like non-volatile strain has been reported, and this is ascribed to the ferroelastic domain switching that constructed a third way for realizing the non-volatile strain state.[87,88] Although much remains to be done for a precise control of the ferroelastic domain switching process, the non-volatile strain associates with it has been proven promising for realizing the non-volatile control over magnetism. For example, S. Zhang and co-authors[195] reported a large and nonvolatile bipolar electric-field-controlled magnetization in $Co_{40}Fe_{40}B_{20}$ on (001)-oriented PMN-PT structure [Fig. 12(a)], which exhibits an electric-field-controlled loop-like magnetization [Fig. 12(b) and (c)]. This effect is related to the combined action of 109 °ferroelastic domain switching and the absence of magnetocrystalline anisotropy in $Co_{40}Fe_{40}B_{20}$. The large and nonvolatile magnetoelectric coupling achieved at room temperature [Fig. 12(d)] is of great importance to achieve electric-controlled magnetic random access memories. For $Co_{60}Fe_{20}B_{20}$ deposited on (011)-oriented PMN-PT, Liu et al.[189] reported up to 90% ferroelastic switching of the polarization corresponding to rotation from an out-of-plane to a purely in-plane direction (71 °and 109 °switching). The high efficient ferroelastic switching produces two distinct, stable, and electrically-reversible lattice strain states, which has been demonstrated to results in a highly energy-efficient, nonvolatile tuning of FMR frequency up to 2.3 GHz in elastically coupled amorphous $Co_{60}Fe_{20}B_{20}$ films.

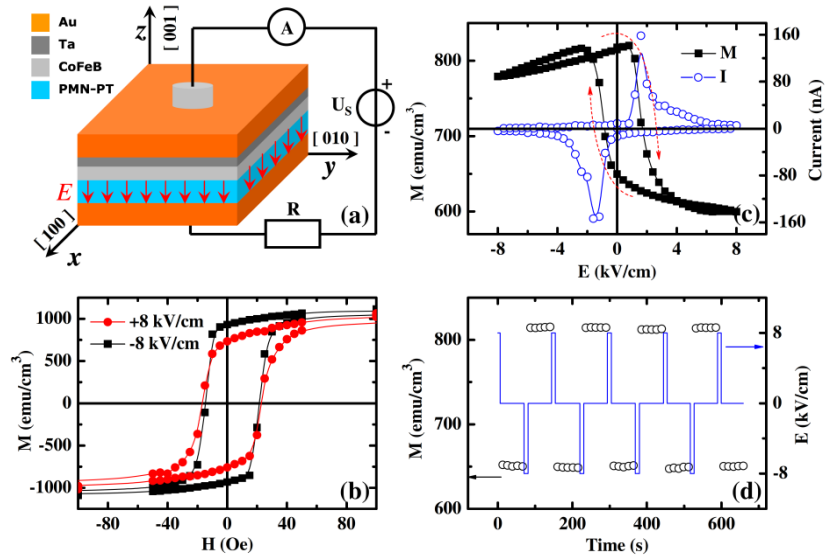

**Fig. 12** (a) Scheme of the sample and experimental configuration. (b) In-plane magnetic hysteresis loops under electric fields of +8 kVcm$^{-1}$ (circle) and -8 kVcm$^{-1}$ (square). (c) Electric-field tuning of the in-plane magnetization (square) and polarization current (open circle) recorded at the same time. (d) The repeatable high/low magnetization states (open circle) switched by pulsed electric fields (blue line).[195]

**3.4 Flexible electronic devices**

In recent years, flexible electronic devices, which offers advantages of being lightweight, foldable, and stretchable, has attracted special attention. Although a great deal of flexible electronic devices was made of organic materials, there has been an increasing trend in the exploration of flexible inorganic devices, such as thin-film transistor,[196-199] magnetoelectronics,[164,200,201] and memories.[202-205] In this part, we merely review the progress on flexible RRAM devices based on oxides.

In RRAM devices, the RS is dominated by the formation and rupture of the CF, as the CF is of small size and perpendicular to the film plane, hence the RS phenomenon can endure moderate strain along the film plane. There have been some reports of flexible memristive memories with a simple cross-point-type array based on various oxide such as $GeO_x$/HfON,[206] aluminum oxide,[207] zinc oxide,[202,208] and $TiO_2$,[209,210] *etc*. These devices show good flexibility, and the performance of RS is not degraded by substrate bending. In order to overcome the cell-to-cell interference issue, Kim *et al.* developed the "NOR" type flexible RRAM with a one transistor-one memristor structure, and performed the first demonstration of random access memory operation of the RRAM on a flexible substrate.[211] Their developed device exhibited reliable and reproducible RS, good endurance and retention properties, along with excellent mechanical stability upon harsh bending. Liang *et al.* has demonstrated ultrathin $WO_3·H_2O$ nanosheets as a promising material to construct a high performance and flexible RRAM device [Fig. 13(a)].[212] The ultrathin $WO_3·H_2O$ nanosheets were exfoliated from the bulk hydrated tungsten oxides ($WO_3·nH_2O$) through wet chemical method. Both of them show excellent RS behavior [Fig. 13(b)], however, the device with $WO_3·H_2O$ nanosheets is superior to both its bulk counterparts and most of the reported typical RS inorganic materials [Fig. 13(c)], possibly due to the abundant vacancy associates in the ultrathin nanosheets. In addition, this ultrathin nanosheets-based RRAM device also shows excellent flexibility [Fig. 13(d)] due to its 2D ultrathin characteristic, which is difficult to obtain in both bulk and nanoparticles.

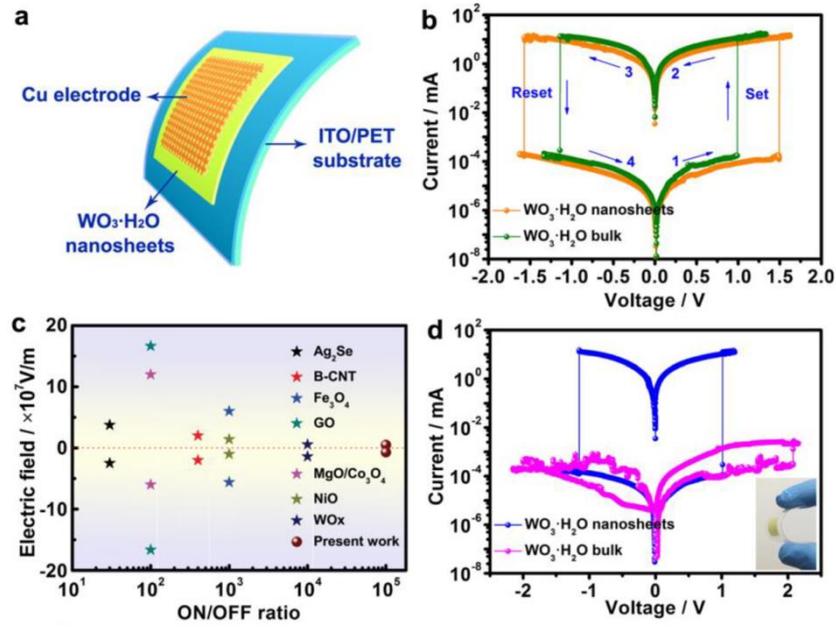

**Fig. 13** (a) Schematic illustration of flexible RRAM device with a Cu/WO$_3$·H$_2$O/ITO-PET configuration. (b) Current versus voltage (I−V) curves of the ultrathin WO$_3$·H$_2$O nanosheets-based and bulk-based RRAM devices at initial cycle. The arrows and numbers indicate the voltage sweep directions. (c) Comparison of the performance of the RRAM device based on ultrathin WO$_3$·H$_2$O nanosheets with other reported typical RS inorganic materials. (d) I−V curves obtained from the WO$_3$·H$_2$O nanosheets-based and bulk-based RRAM devices after 2000 and 100 bending tests, respectively. Inset is a digital photograph of a bent RRAM device.[212]

## 4. Magnetic field effect

The study of magnetic field on the physical properties of oxide thin films is an important issue in condensed matter physics, which also shows temping prospect of applications in spintronic devices and magnetic sensors. The discovery of colossal magnetoresistance (CMR) effect, as well as the investigation of full oxide magnetic tunnel junctions (MTJ) has greatly prompt the study of magnetic field on thin film oxides. The effect of magnetic field on the electrical properties of oxide thin films has been studied in order to combine charge-tuning devices with spintronics devices.

**4.1 Magnetoresistance effect**

In perovskite manganites, applying magnetic field usually facilitates double exchange interaction which favors metallic conductivity, inducing the well-known CMR effect.[213] The CMR effect has been widely studied in manganite,[26,28,29,214-216] and it always shows a peak value near the phase boundary, which reflects the fact that the phase competition is susceptible to external magnetic field.[27] In recent years, it has been found that despite the near cubic structure of perovskite manganites, the electrical transportation behavior actually depends strongly on the magnetic field direction, resulting in very large AMR effect in manganite single crystals[217,218] and thin films.[219-223] The AMR effect shows different magnetic-field and temperature dependence from that in 3$d$ ferromagnetic alloys, suggesting different mechanisms behind.[217,224,225] The AMR effect in manganite thin films can be tuned by epitaxial strain and has attracted lots of attentions.[221,226-229]

**4.2 Magnetic coupled RRAM devices**

The magnetic field couples with RRAM devices in two different ways. One is by the effective tuning of the resistance switching process, and the other is by realizing more resistance states through magnetoresistance effect. Thakare *et al.*[230] reported a bipolar RRAM of CFO/LSMO which can be gated with high sensitivity by low magnetic fields. The switching shows a strong CFO thickness dependence only under applied magnetic field, suggesting an effective coupling with the interfacial magnetoresistive LSMO layer. Li *et al.* demonstrated three-state RS in films combining multiferroic CFO/PZT bilayer with semiconductor ZnO layer.[231] The three states are distinguished by various stimuli combination of electric voltage pulse and magnetic bias, and are attributed to changes in the charge carrier states modulated by magnetoelectric coupling between CFO and PZT layers, as well as interface polarization coupling between PZT and ZnO layers. In a Si-SiO$_2$-MgO device, the nonvolatile RS effect can be controlled by applying a magnetic field.[232] The magnetic field will delay the transition from HRS to LRS, and could even suppress this transition. Das *et al.*[233] observed sharp switching between low and high conducting states having a large ratio of currents (~10$^7$) in NiO film. Besides the switching effect, an extraordinary large magnetoconductance (~98% at 20 K) has been found in the device, signifying multifunctional applications.

The RS effect can be utilized as a simple and controllable way to produce nanoscale devices, such as an artificial MTJ.[234] The spin transportation in memristive devices has also been reported in a number of MTJs,[235-238] it is found that applying an electric field may induce stable, reversible two-resistance states with substantial magnetoresistance change. Li *et al*. reported the use of composite barrier layers of CoO-ZnO to fabricate the spin memristive Co/CoO-ZnO/Co MTJ.[239] They found a bipolar resistance switching ratio high up to 90 [Fig. 14(a)], and the tunnel magnetoresistance (TMR) ratio of the high resistance state reaches 8% at room temperature [Fig. 14(b)]. At low resistance state, the TMR ratio vanishes [inset of Fig. 14(b)]. The bipolar resistance switching is explained by the MIT of CoO$_{1-x}$ layer due to the migration of oxygen ions between CoO$_{1-x}$ and ZnO$_{1-x}$ [Fig. 14(c)]. These novel phenomena enable integrating different functionalities into a single device, which has potential applications including multi-state storage devices and logic circuit.[238]

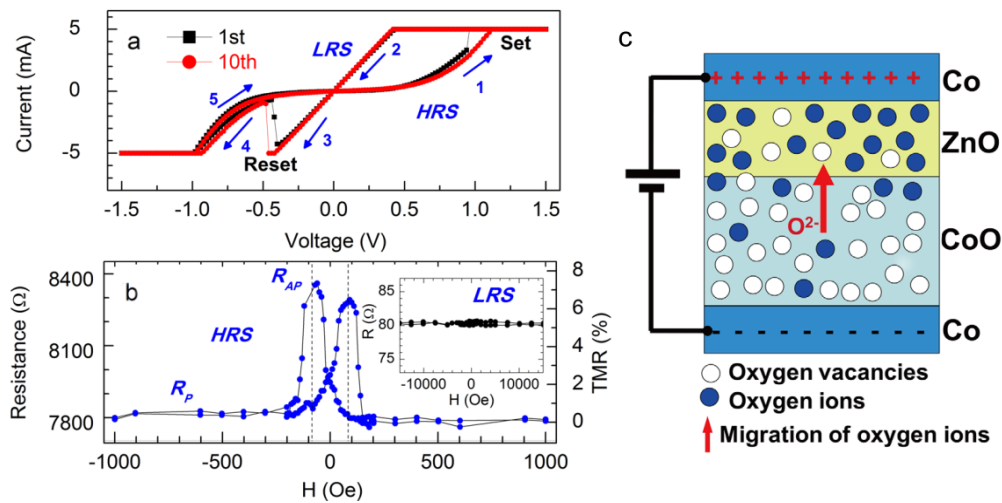

**Fig. 14** (a) The I-V characteristic of Ag(30 nm)/Co(10 nm)/CoO-ZnO(2 nm)/Co(30 nm)/Ag(60 nm) junction with area 0.1 mm 3 0.1 mm, and (b) the tunneling magnetoresistance of the junction. The inset in (b) shows the R-H curve of the low resistance state. Schematics of the migration of oxygen ions between very thin CoO and ZnO layers under a positive voltage, and resulting MIT of $CoO_{1-x}$ in Co/CoO-ZnO/Co junctions.[239]

**4.3 Magnetoelectric and magnetocapacitance effects**

As has been mentioned above, the interplay between electric polarization and magnetization in multiferroic materials is very important for both fundamental physics and spintronic devices. Manipulation of magnetism by electric polarization has been manifested both in single-phase multiferroics and composites. Conversely, the possibility of electric polarization reversal induced by magnetic fields has been reported in multiferroics. Keeney *et al*. demonstrated switching of ferroelectric polarization by magnetic fields at room temperature in multiferroic Aurivillius phase thin films.[240] For practical applications, the weak ME coupling in single-phase multiferroics is always a big challenge.[241] Lu *et al*. reported a remarkable magnetically induced electric polarization in $DyMnO_3$ thin films, and they found an enhancement of electric polarization up to 800% upon 2 Tesla at 2 K.[42]

An efficient coupling between magnetism and dielectric properties results in phenomenon such as magnetocapacitance effect. This phenomenon has been studied both in single phase materials[47,242-246] and composite systems.[247-250] For example, the magnetocapacitance effect has been found in perovskite heterostructures such as LCMO/BTO[247] and LCMO/$Ba_{1-x}Sr_xTiO_3$[251] superlattices, manifesting an effective ME coupling in these structures. Colossal magnetocapacitance has been reported in electronic phase separated manganite thin films, and the competition of microscopic ferromagnetic metallic and charge ordered insulating clusters is believed to play a major role in the exhibited colossal magnetocapacitance effect.[245]

**5. Photoinduced effect**

Photo-coupled phenomena have attracted wide interest due to the special advantages of light, such as high speed, low power consumption and green energy, etc. In thin film oxides, photoinduced effect such as photostrictive effect,[77] photoinduced electronic phase transition,[252] photovoltaic effect,[253] and photoinduced spin dynamics,[254] indicates a diversity of degrees of freedoms that couples effectively with light.

**5.1 Photo-coupled RRAM devices**

Light-controlled RS memory has photon-charge bifunction. The resistance state can be modified not only by voltage pulses, but also by means of light. M. Ungureanu et al.[255] reported a new type of RS memory device with a simple metal/$Al_2O_3$/$SiO_2$/Si structure. In order to verify the performance of the devices when used as core of a binary memory cell, they observe that data can be accurately written, read, and erased only under illumination [Fig. 15B] by applying 5 ms voltage pulses in the sequence −10 V/+6 V/ +10 V/+6 V both in dark conditions, as well as under illumination with UV (390 nm) and IR (950 nm) light emitting diodes (LED) with irradiances up to 2.5 mW/cm$^2$.

The change in the remnant current was interpreted as a modulation of the trapped electrons in the $Al_2O_3$ layer. Upon the introduction of electrons from the Si layer, the resistivity of this charge storage layer increases. For positive bias, a depletion region is generated in Si at the interface with the $SiO_2$/$Al_2O_3$. The photogenerated electrons are injected in the $Al_2O_3$ layer when the electric field is high enough to overcome this energy barrier [Fig. 15A 1, 2 and 3]. At negative voltages, the charge will be removed gradually [Fig. 15A 4, 5 and 6]. When the device is measured in dark conditions, the modulation in the remnant current is strongly suppressed, since the source of electrons is only the small dark current [Fig. 15A (a)]. The memory devices were tested and support more than $10^4$ write/read/erase/read cycles, with a data retention time, obtained by extrapolation, of about one year. Moreover, Light-Controlled RS behavior has also been found in ZnO NRs/Nb:STO Single Crystal Diodes[256] and ITO/ZnO NR/Au junction[257].

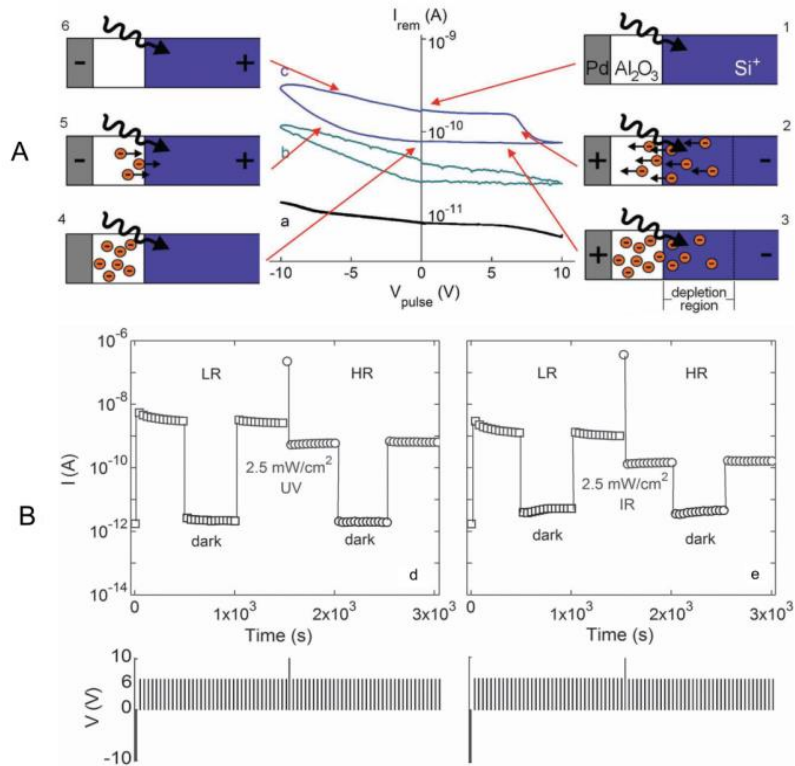

**Fig. 15** A. Remnant current hysteresis switching loops. The measurement was taken under different UV illumination conditions. For the 2.5 mW/cm$^2$ irradiance case, the behavior was described by the trapped charges. B. Data retention capability in dark conditions and under illumination with UV light (a), and IR light (b) in a metal/Al$_2$O$_3$/SiO$_2$/Si structure. The voltage-pulses applied are displayed in the lower part of the figures.[255]

Recently, a multifunctional optoelectronic RS memory has been demonstrated in a simple ITO/CeO$_{2-x}$/AlO$_y$/Al structure [Fig. 16(a)].[258] As a result of the photo-induced detrapping, electrode-injection and retrapping of electrons in the CeO$_{2-x}$/AlO$_y$/Al interfacial region, the device can be optically programmed and electrically erased. The device shows broadband, linear and persistent photoresponses, which has been utilized in integration of the demodulating [Fig. 16(b)], arithmetic [Fig. 16(c)] and memory [Fig. 16(d)] functions in a single device. This multi-functional device could be useful for future optoelectronic interconnect systems.

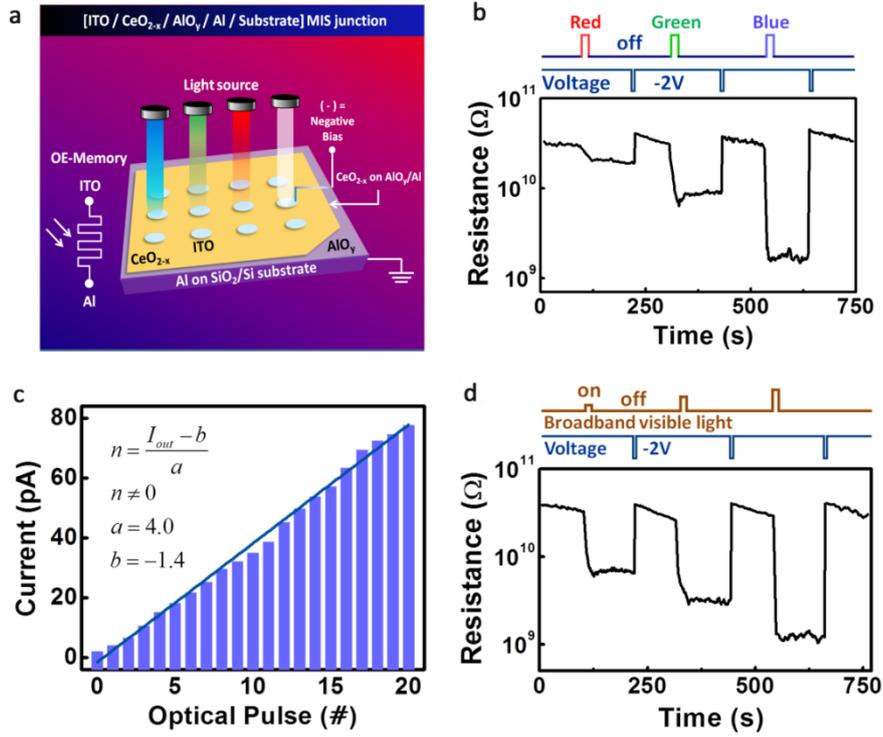

**Fig. 16** (a) Schematic illustration of the ITO/CeO$_{2-x}$/AlO$_y$/Al junction structure and operation principle of the multifunctional optoelectronic RS memory (OE-Memory). (b) Wavelength-dependent photoresponse of the device upon being exposed to the monochromic illumination with the wavelength of 638 nm, 560 nm and 499 nm and identical intensities of 6 pW/μm$^2$, respectively, enables the device as a optoelectronic demodulator. (c) Stepwise and linear relationship between the device output current and the number of the input light pulses, which form the basis for application of computing functions. The read voltage is 0.1 V. $I_{out}$, $n$, $a$ and $b$ stand for the output current (pA) of the device, number of the light pulse inputs, photocurrent ramping step of ~ 4.0 (pA) per light illumination, and a constant of ~ -1.4 (pA), respectviely, for n≠0. The dark current is ~ 2 pA for n=0. (d) Intensity-dependent photoresponse of the device under broadband illumination with the intensities of 8 pW/μm$^2$, 21 pW/μm$^2$ and 60 pW/μm$^2$, respectively, renders promising multilevel storage capabilities.[258]

## 5.2 Photoinduced phase transitions

Photoinduced phase transitions have attracted special interest because they can change the properties of complex materials on the ultrafast timescale.[259-262] In perovskite manganites, photoinduced insulator-to-metal transition has been frequently reported,[263-268] and the photo illumination has been found to coupled coherently in these materials, resulting in modifications of a variety of the physical properties due to the strong coupling between spin, charge, orbital and lattice. In PCSMO thin films, both photoinduced metal-to-insulator transition and photoinduced insulator-to-metal transition, have been confirmed.[269] This is a demonstration of photons as an effective way in overcoming the large potential barrier due to the long-range elastic energy between the insulator and metallic phase. Possibly due to the high energies associated with photons, the photoinduced state could differ greatly from states produced with stimuli close to equilibrium. For example, in a charge-orbital ordered $Nd_{0.5}Sr_{0.5}MnO_3$ thin film, the photoinduced state was found to be structurally ordered, homogeneous, metastable insulating state with crystallographic parameters different from any thermodynamically accessible state.[270]

In vanadium dioxide ($VO_2$), a first-order transition between a high-temperature metallic phase and a low-temperature semiconducting phase has been evidenced.[130] A nonthermal driven ultrafast phase transition can be achieved under photo illumination.[271] The transition takes place near room temperature, and has been heavily studied for both fundamental physics and possible applications such as switching devices.[58,272] There has been hot debate regarding the driving force of this insulator-metal transition, and contributions of both electron-lattice interactions and electron-electron interactions have been proposed.[273,274] Advanced time-resolved methods have been developed to capture the phase transition processes, and much has been gained about the details during the transition processes.[274,275] For instance, Morrison *et al.* utilized ultrafast electron diffraction to monitor the crystal structure of $VO_2$ and simultaneously measured its optical properties to monitor the electronic state, and they have found an interesting metastable state that preserves the initial semiconducting crystal structure while shows metal-like properties.[275]

## 5.3 Photovoltaic effect in ferroelectric thin films

Another important consequence of photo illumination is the photovoltaic effect, which has been frequently studied in ferroelectric thin films. The ferroelectric-photovoltaic device, in which a homogeneous ferroelectric material is used as a light absorbing layer, has been investigated with numerous ferroelectric oxides.[276-278] The ferroelectric-photovoltaic effect is distinctly different from the typical photovoltaic effect in semiconductor *p-n* junctions in that the polarization electric field is the driving force for the photocurrent in ferroelectric-photovoltaic devices.[279,280] In addition, the anomalous photovoltaic effect, in which the voltage output along the polarization direction can be significantly larger than the bandgap of the ferroelectric materials, has been frequently observed in ferroelectric-photovoltaic devices.[253,281] For most ferroelectric materials, the relatively wide band gap limited the photovoltaic efficiency, and utilization of narrow-band-gap ferroelectrics is therefore a promising route towards their application in both the novel optoelectronic and the solar energy devices.[282,283] Multiferroic material BFO, which possesses a narrow band gap (~2.2 eV) offers an exciting opportunity for such applications. A significant photovoltaic effect has already been observed in single crystalline BFO in both bulk[279,284,285] and thin film forms,[286-290] but still needs improving.[253,291-293]

Yang *et al*.[253] studied the photovoltaic effect on the BFO film with ordered domain strips and lateral device configuration [Fig. 17(b)]. They observed that the photovoltage in the BFO film increased linearly with the total number of domain walls (DW) along the net polarization direction (perpendicular to the domain walls, poled state), and the photovoltaic effect vanished along the direction perpendicular to the net polarization direction (as grown state) [Fig. 17(a)]. The intrinsic potential drop at domain walls (~10 mV), arising from the component of the polarization perpendicular to the domain wall, induces a huge electric field of ~5 $\times 10^6$ Vm$^{-1}$ in the narrow domain wall, which was suggested to be the driving force for the dissociation of the photo-generated exciton. The illuminated domain walls act as nanoscale photo-voltage generators connected in series, wherein the generated photocurrent is continuous and the photo-generated voltage accumulates along the net polarization direction.

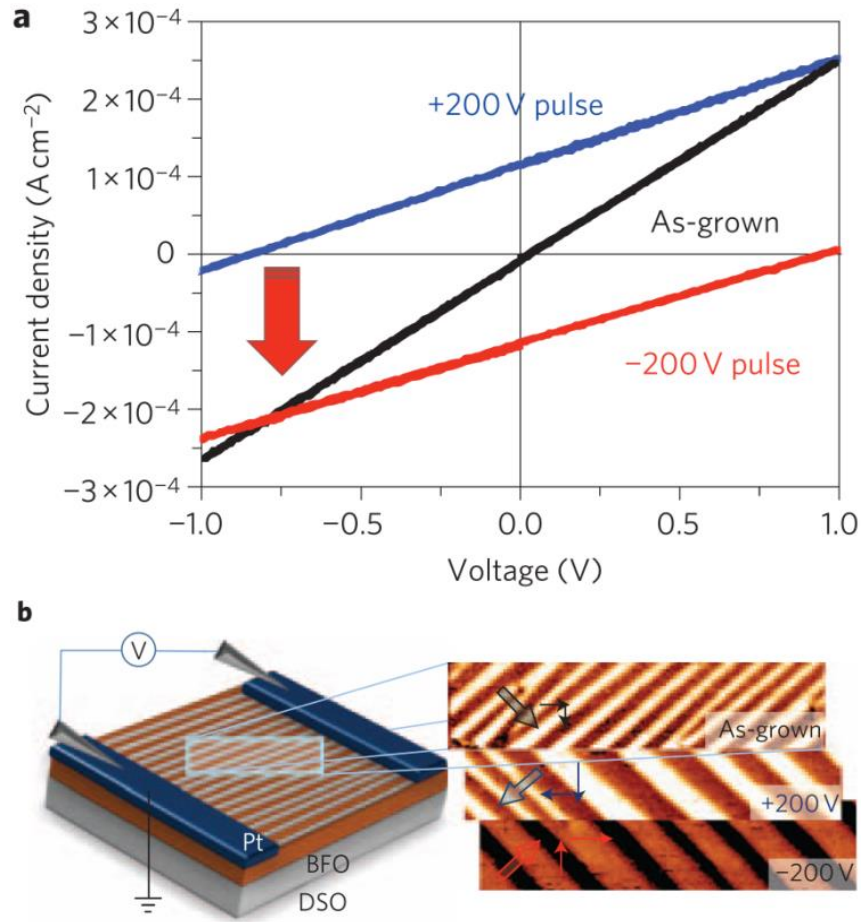

**Fig. 17** (a) Light I–V measurement in the DW$_\parallel$ geometry shows no observable photovoltaic effect, as grown. On rotation of the domain structure to the DW$_\perp$ configuration after application of +200 V voltage pulses to the in-plane device structure, a photovoltaic effect is observed. (b) Corresponding PFM images of the as-grown (top panel), 200 V poled (middle panel), and -200 V poled (bottom panel) device structures. The arrows indicate the in-plane projection of the polarization and the net polarization direction for the entire device structure.[253]

In the ferroelectric based photovoltaic devices, it is suggested that the depolarization field may be the dominating driving force for the separation of photo-generated charge carrier-pairs. This depolarization field should be closely related to the degree of screening of the spontaneous polarization, which depends on both the properties of ferroelectric materials and electrode materials. Chen *et al.*[292]

studied the anomalous photovoltaic effect in the device with a structure of Au/BFO/Au and found that the photocurrent output was increased 25 times when the Au electrode was replaced by ITO. The great enhancement of photocurrent was attributed to the increased depolarizing field.

Furthermore, the photovoltaic effect in ferroelectric films has also provided a new method to sense the polarization direction non-destructively in ferroelectric memory. For example, Guo *et al.* reported a novel approach to create a non-volatile memory based on the ferroelectric polarization-dependent photovoltaic effect.[294] Recently, Wang *et al.* presented a new non-destructive readout by using photo-recovered surface potential contrast.[295] By introducing laser illumination, they found that the surface potential contrast between the adjacent oppositely polarized domains can be recovered, as has been revealed by kelvin probe force microscopy (KPFM) measurement [Fig. 18(e) and 18(f)]. The mechanism associates with the redistribution of the photoinduced charges driven by the internal electric field. The have created a 12-cell memory pattern based on BFO films to show the feasibility of such photo-assisted non-volatile and non-destructive readout of the ferroelectric memory [Fig. 18 (a)-(d)].

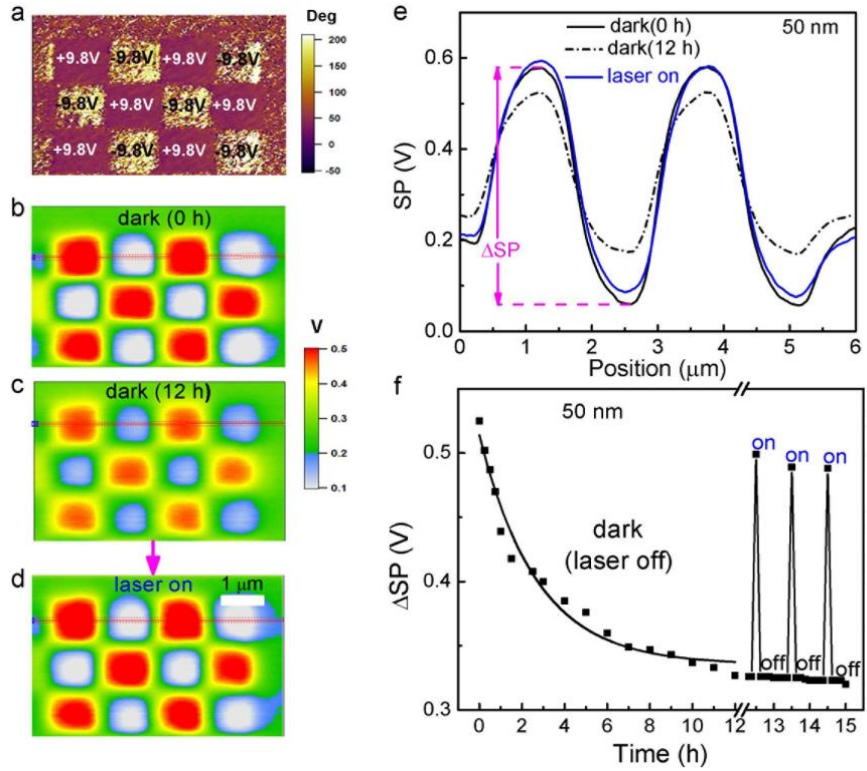

**Fig. 18** (a) Out-of-plane PFM phase image obtained after poling for 50-nm-thick BFO films with polarized pattern. (b)-(d) KPFM images measured without (dark) and with light illumination by a 375 nm laser (power density of 40 mWcm$^{-2}$). (e) Surface potential profiles obtained from (b)–(d), respectively. (f) Time dependence of DSP. ''on'' and ''off'' represent laser on and laser off, respectively.[295]

## 6. Conclusions and outlooks

Oxide thin films exhibit interesting and sometimes unexpected property changes upon external stimulus. These novel phenomena provide the prototype for the next generation of devices. Although many progresses have been made in the systems, there are still many challenges and opportunities. As an example, the emergent phenomenon at oxide interface is an interesting yet not well explored subject.[62,66,73] In multiferroic based devices, a central idea is the electric control of magnetism that would enable scalable and green energy spintronic devices[296]. Although the electric induced energy-efficient control of a spin-valve device at room temperature has been demonstrated in BFO based heterostructures,[99] there is challenge in improving reliability of the devices, possibly due to oxidation of the ferromagnetic metal at the interface under the large electric field. In this sense, study of full oxide spintronics would be of importance to make the device robust.[98,297] In RRAM based devices, the study of ion transportation have brought about new direction toward physics in nanoscale, and it could be useful in designing novel electronic devices with nanoscale sizes. Difficulties lie in effective characterizations of the ionic channels, and an effective way for improving the stability and uniformity of the nanoscale devices. Given the strong composition-structure-property relationship in many complex oxides, thin film oxides show sizable and sometimes unexpected response to external stimuli, making them a wonderful playground for studying fundamental physics and exploring novel devices. In this case, in-situ structural and chemical characterizations during film growth are desirable.